\documentclass[
copyright,creativecommons]{eptcs}
\providecommand{\event}{MFPS 2021}
\usepackage{graphicx}
\providecommand{\event}{MFPS 2021}

\providecommand{\urlalt}[2]{\href{#1}{#2}}
\providecommand{\doi}[1]{doi:\urlalt{http://dx.doi.org/#1}{#1}}

\usepackage[utf8]{inputenc}
\usepackage[english]{babel}
\usepackage{amsmath, amsthm}
\makeatletter
\newcommand\Label[1]{\refstepcounter{equation}(\theequation)\ltx@label{#1}}
\makeatother
\usepackage{amssymb}
\usepackage{stmaryrd}
\usepackage{graphicx}
\usepackage{hyperref}

\hypersetup{
    colorlinks=true,
    linkcolor=blue,
    filecolor=magenta,      
    urlcolor=cyan,
}

\usepackage[all]{hypcap}
\usepackage{proof}

\usepackage[disable]{todonotes}

\makeatletter
\newcommand\etc{etc\@ifnextchar.{}{.\@}\xspace}

\makeatother

\newcommand{\intf}[1]{\left\llbracket #1 \right\rrbracket}

\renewcommand{\int}[1]{\left[#1\right]}

\newcommand{\res}[1]{\overline{#1}}

\newcommand{\lcs}[1]{\res{\intf{#1}^{\circ}}}
\newcommand{\lc}[1]{\res{#1^{\circ}}}

\newcommand{\alt}{~\mid~}

\newcommand{\iso}{\leftrightarrow}

\newcommand{\inl}[1]{\mathtt{inj}_l{\;#1}}
\newcommand{\inr}[1]{\mathtt{inj}_r{\;#1}}

\newcommand{\pv}[2]{\langle #1,#2 \rangle}

\newcommand{\clause}[2]{\mid ~ #1 \iso #2}
\newcommand{\clauses}[1]{\left\{ ~#1~ \right\}}
\newcommand{\isobasique}{\clauses{\clause{v_1}{v_1'}\mid\dots\clause{v_n}{v_n'}}}
\newcommand{\isobasiqueinv}{\clauses{\clause{v_1'}{v_1}\mid\dots\clause{v_n'}{v_n}}}

\newcommand{\isoterm}{\omega}

\newcommand{\entailval}{\vdash_v}
\newcommand{\entailiso}{\vdash_{\isoterm}}

\usepackage{color}
\def\bR{\begin{color}{red}}
\def\bB{\begin{color}{blue}}
\def\bM{\begin{color}{magenta}}
\def\bC{\begin{color}{cyan}}
\def\bW{\begin{color}{white}}
\def\bBl{\begin{color}{black}}
\def\bG{\begin{color}{green}}
\def\bY{\begin{color}{yellow}}
\def\e{\end{color}\xspace}
\newcommand{\bit}{\begin{itemize}}
\newcommand{\eit}{\end{itemize}\par\noindent}
\newcommand{\ben}{\begin{enumerate}}
\newcommand{\een}{\end{enumerate}\par\noindent}
\newcommand{\beq}{\begin{equation}}
\newcommand{\eeq}{\end{equation}\par\noindent}
\newcommand{\beqa}{\begin{eqnarray*}}
\newcommand{\eeqa}{\end{eqnarray*}\par\noindent}
\newcommand{\beqn}{\begin{eqnarray}}
\newcommand{\eeqn}{\end{eqnarray}\par\noindent}

\usepackage{keycommand}

\newtheorem{theorem}{Theorem}[section]
\newtheorem{corollary}[theorem]{Corollary}
\newtheorem{lemma}[theorem]{Lemma}

\newtheorem{property}[theorem]{Property}
\newtheorem{definition}[theorem]{Definition}
\newtheorem{example}[theorem]{Example}

\newtheorem{example*}[theorem]{Example*}
\newtheorem{examples*}[theorem]{Examples*}

\newtheorem{remark}[theorem]{Remark}
\newtheorem{remark*}[theorem]{Remark*}

\title{Categorical Semantics of Reversible Pattern-Matching}

\def\titlerunning{Categorical Semantics of Reversible Pattern-Matching}

\author{Kostia Chardonnet \institute{LMF, Univ. Paris Saclay.}\institute{IRIF, Univ. Paris.}\email{kostia@lri.fr} \and Louis Lemonnier\institute{LMF, Univ. Paris Saclay.}\email{lemonnier@lsv.fr} \and Benoît Valiron\institute{LMF, CentraleSupélec, Univ. Paris Saclay}\email{benoit.valiron@universite-paris-saclay.fr}}

\def\authorrunning{Kostia Chardonnet, Louis Lemonnier, Benoît Valiron}

\begin{document}
\maketitle 

\begin{abstract} 
  This paper is concerned with categorical structures for reversible
  computation. In particular, we focus on a typed, functional
  reversible language based on Theseus. We discuss how join inverse rig
  categories do not in general capture pattern-matching,
  the core construct Theseus uses to enforce reversibility.  We then
  derive a categorical structure to add to join inverse rig
  categories in order
  to capture pattern-matching. We show how such a structure makes an
  adequate model for reversible pattern-matching.
\end{abstract}

\section{Introduction}

In this paper, we are concerned with the semantics of reversible
programming languages.
The idea of reversible computation comes from Landauer and
Bennett~\cite{Landauer61,bennett1973logical} with the analysis of its
expressivity, and the relationship between irreversible computing and
dissipation of energy.
This lead to an interest in reversible computation~\cite{bennett2000notes,aman2020foundations}, both with a
low-level
approach~\cite{caroe2012design,wille2016syrec,saeedi2013synthesis},
and from a high-level
perspective~\cite{lutz1986janus,yokoyama2007reversible,yokoyama2016fundamentals,james2012information,%
james2014theseus,sabry2018symmetric,yokoyama2011reversible,thomsen2015interpretation,JacobsenKT18}.
Reversible programming lies on the latter side of the spectrum, and
two main approaches have been followed.  Embodied by
Janus~\cite{lutz1986janus,yokoyama2007reversible,yokoyama2010reversible,yokoyama2016fundamentals}
and later R-CORE and R-WHILE~\cite{gluck2019reversible}, the first one
focuses on imperative languages whose control flow is
inherently reversible ---the main issue with this aspect being tests
and loops. The other approach is concerned with the design of
functional languages with structured data and related
case-analysis, or \emph{pattern-matching}~\cite{yokoyama2011reversible,thomsen2015interpretation,james2014theseus,sabry2018symmetric,JacobsenKT18}. To
ensure reversibility, strong constraints have to be established on the
pattern-matching in order to maintain reversibility.

In general, reversible computation captures \emph{partial injective
  maps}~\cite{gluck2019reversible} from inputs to outputs, or,
equivalently in this paper, \emph{partial isomorphisms}. Indeed, from
a computational perspective reversibility is understood as a
time-local property: if each time-step of the execution of the
computation can soundly be reversed, there is no overall condition on
the global behavior of the computation. In particular, this does not
say anything about termination: a computation seen as a map from
inputs to outputs might very well be partial, as some inputs may
trigger a (global) non-terminating behavior.

The categorical analysis of partial injective maps have been
thoroughly analyzed since 1979, first by
Kastl~\cite{kastl1979inverse}, and then by Cockett and
Lack~\cite{cockett2002restriction-I,cockett2003restriction-II,%
cockett2007restriction-III}.  This led to the development of
\emph{inverse category}: a category equipped with an inverse operator
in which all morphisms have partial inverses and are therefore
reversible. The main aspect of this line of research is that
partiality can have a purely algebraic description: one can introduce
a restriction operator on morphisms, associating to a morphism a
partial identity on its domain. 

This categorical framework has recently been put to use to develop
the semantics of specific reversible programming constructs and concrete
reversible languages: analysis of recursion in the context of
reversibility~\cite{axelsen2016join,kaarsgaard2019inversion,kaarsgaard2019engarde},
formalization of reversible
flowchart languages~\cite{gluck2017categorical,kaarsgaard2019condition},
analysis of
side-effects~\cite{heunen2015reversible,heunen2018reversible},
\textit{etc}. Interestingly enough however, the adequacy of the
developed categorical constructs with
reversible \emph{functional} programming languages has been seldom
studied. For instance, if Kaarsgaard \textit{et
  al.}~\cite{kaarsgaard2017join} mention Theseus as a potential
use-case, they do not discuss it in details. So far, the semantics of
functional and applicative reversible languages has always been done
in \emph{concrete} categories of partial
isomorphisms~\cite{kaarsgaard2019engarde,kaarsgaard2021join}.

In particular, one important aspect that has not been addressed yet in
detail is the categorical interpretation of pattern-matching. If
pattern-matching can be added to reversible imperative
languages~\cite{gluck2019reversible}, it is particularly relevant in
the context of functional languages where it is one of the core
construct needed for manipulating structured data. This is for
instance emphasized by
the several existing languages making use of
it~\cite{yokoyama2011reversible,thomsen2015interpretation,james2014theseus,JacobsenKT18,sabry2018symmetric,chardonnet2020curry}.
In the literature, pattern-matching has either been considered in the
context of a Set-based semantics~\cite{gluck2019reversible}, or more
generally in categorical models making heavy use of rig
structures~\cite{carette2016computing} or
co-products~\cite{kaarsgaard2019engarde,kaarsgaard2021join} to
represent it. If such rich structures are clearly enough to capture
pattern-matching, we claim that they are too coarse, and that a weaker
structure is enough for characterizing pattern-matching.

\paragraph{Contributions.}
In this paper, we make a proposal for a general categorical
interpretation of pattern-matching in the context of inverse categories,
without having to rely on external structure such as coproducts.  More
specifically, we study the categorical semantics of a typed, linear and
reversible language in the style of
Theseus~\cite{james2014theseus}. We develop and discuss
\emph{pattern-matching categories}: 
a categorical construction shown to be sufficient
to model the pattern-matching at the
heart of the operational semantics of the language. In particular,
in this category we can mimic the notion of clauses and
values to be matched against them, without to have to rely 
on coproducts.
We conclude the paper with a proof that
pattern-matching categories make adequate
models for the considered language, thus
confirming the validity of the approach.

This paper is organized as follows: first, in
Section~\ref{sec:language} we present the language based
on~\cite{sabry2018symmetric}, its type system and rewriting system
along with the usual safety properties.
We then give some reminder on inverse
categories and set the theoretical background for the rest of the
paper in Section~\ref{sec:inv_cat}. Basic knowledge in category theory
is assumed. Finally, sections~\ref{sec:sem_iso} and~\ref{sec:pattern} are
focused on the categorical semantics of the language. The proofs are
available in the appendix.

\section{The language}
\label{sec:language}

\newcommand{\setden}[1]{{\{\!|{#1}|\!\}}}

In this section, we present a small, formal reversible functional
programming language in the like of Theseus~\cite{james2014theseus}
and its later
developments~\cite{sabry2018symmetric,chardonnet2020curry}. The core
feature of Theseus is to define reversible control-flow using
pattern-matching. Doing so elegantly bridges functional programming
and reversible computation in a typed manner: well-typed, terminating
programs describe isomorphisms between types. In previous
---untyped--- approaches~\cite{yokoyama2011reversible},
pattern-matching was simply considered as one language feature among
many, without any special attention devoted to it, and the reversible
aspect was taken care of separately. On the contrary, Theseus uses
pattern matching as the core feature to make the language reversible.

In the context of a typed pattern-matching, the compiler can easily
verify two crucial properties: (i) non-overlapping patterns in
clauses, ensuring deterministic behavior, and (ii) exhaustive coverage
of a datatype with the patterns, ensuring
totality. In~\cite{sabry2018symmetric}, these properties are shown
sufficient to produce a simple first-order reversible programming
language.
In this paper we relax the constraints on exhaustivity, making it
possible to define partial functions.

\subsection{Terms and Types}
\label{sec:1st-order-finite}

The language we focus on in this paper is a weakened version of the
language presented in~\cite[Sec.2]{sabry2018symmetric}. In particular,
we allow non-exhaustive pattern-matching and enum types.
The language is typed and consists of two layers: values and functions
(called ``isos'' in~\cite{sabry2018symmetric}). It is parametrized by
a set of enum types (spanned by $\alpha$) and their constant values
(spanned by $c_\alpha$). The language whose only type $\alpha$ is the
unit-type with one single unit constant will be called the
\emph{minimal language}.
\begin{alignat*}{10}
  &\text{(Value types)}\quad & a, b &~~&&::= ~&\quad& \alpha \alt a \oplus
  b \alt a \otimes b\\
&\text{(Iso types)} & T &&&::=&& a \iso b \\[1ex]
&\text{(Values)} & v &&&::=&& c_\alpha \alt x \alt \inl{v} \alt \inr{v} \alt
                            \pv{v_1}{v_2}\\
&\text{(Functions)} & \isoterm &&&::=&& 
           \isobasique
           \\
&\text{(Terms)} & t &&& ::= && v \alt \isoterm\,t
\end{alignat*}

The language comes with two kinds of judgments, one for terms and one
for functions (or \emph{isos}).
We denote typing contexts as $\Delta$, they stand
for sets
of typed variables $x_1:a_1,\ldots, x_n:a_n$.
We then write $\Delta \entailval t : a$ for a well-typed term
and $\entailiso \isoterm : a \iso b$ for a well-typed
function. 

Beside the linearity aspect (used to ensure injectivity), the typing
rules for terms are standard. The typing judgment of a term is valid
if it can be derived from the following rules. 
\[\begin{array}{c}
\infer{
  \entailval c_\alpha : \alpha,
}{}
\qquad
\infer{
  x:a\entailval x:a,
}{}
\qquad
\infer{
  \Delta_1,\Delta_2\entailval\pv{v_1}{v_2} : a\otimes b.
}{
  \Delta_1\entailval v_1 : a
  &
  \Delta_2\entailval v_2 : b
}
\\[1.5ex]
\infer{
  \Delta\entailval\inl{v} : a\oplus b,
}{
  \Delta\entailval v : a
}
\qquad
\infer{
  \Delta\entailval\inr{v} : a\oplus b,
}{
  \Delta\entailval v : b
}
\qquad
\infer{
  \entailval \isoterm~t : b
}{
  \entailval t : a & \entailiso \isoterm : a \iso b
}
  \end{array}
\]
In particular, while a value can have free variables, a term is always
closed. This difference is explained by the fact that values are used
as patterns in clauses. The typing rule for isos is as follows.
\begin{equation}\label{eq:typ-iso-specialized}
\infer{ 
  \entailiso 
  \isobasique : a \iso b,
}{
  \begin{array}{@{}l@{}}
    \Delta_1\entailval v_1 : a 
    \\
    \Delta_1\entailval v'_1 : b
  \end{array}
  &
  \ldots 
  &
  \begin{array}{@{}l@{}}
    \Delta_n\entailval v_n : a 
    \\
    \Delta_n\entailval v'_n : b
  \end{array}
  &
  \begin{array}{@{}l@{}}
    \forall i\neq j, v_i\bot v_j
    \\
    \forall i\neq j, v'_i\bot v'_j
  \end{array}
}
\end{equation}
The rule relies on the condition that both left- and
right-hand-side patterns in clauses are orthogonal, enforcing
non-overlapping. The rules for deriving orthogonality of values (or
patterns) are the following.
\[
  \begin{array}{c}
    \infer{c_\alpha~\bot~d_\alpha}{c_\alpha\neq d_\alpha}
    \qquad
    \infer{\inl{v_1}~\bot~\inr{v_2}}{}
    \qquad
    \infer{\inr{v_1}~\bot~\inl{v_2}}{}
    \\[1.5ex]
    \infer{\inl{v_1}~\bot~\inl{v_2}}{v_1~\bot~v_2}
    \quad
    \infer{\inr{v_1}~\bot~\inr{v_2}}{v_1~\bot~v_2}
    \quad
    \infer{\pv{v}{v_1}~\bot~\pv{v'}{v_2}}{v_1~\bot~v_2}
    \quad
    \infer{\pv{v_1}{v}~\bot~\pv{v_2}{v'}}{v_1~\bot~v_2}
  \end{array}
\]
\noindent Left and right injections generates disjoint subsets of
values, and distinct constants of an enum type $\alpha$ are
orthogonal.

\subsection{Operational semantics} 
\label{sec:iso-iso}

The language is equipped with a simple call-by-value operational
semantics on terms based on matching and substitution. We recall the
formalization proposed in~\cite{sabry2018symmetric}, with the notion
of valuation: partial map from a finite set
of variables (the support) to a set of values. We denote the matching
of a value $w$ against a pattern $v$ and its associated valuation
$\sigma$ as $\sigma[v] = w$. It is defined as follows.
\[
\infer{\sigma[c_\alpha] = c_\alpha}{}
\quad
\infer{\sigma[x] = v}{\sigma = \{ x \mapsto v\}}
\quad
\infer{\sigma[\inl{v}] = \inl{w}}{\sigma[v] = w}
\quad
\infer{\sigma[\inr{v}] = \inr{w}}{\sigma[v] = w}
\]

\[
  \infer{
  \sigma[\pv{v_1}{v_2}] = \pv{w_1}{w_2}
}{
  \sigma_2[v_1] = w_1
  &
  \sigma_1[v_2] = w_2
  &
  \text{supp}(\sigma_1) \cap \text{supp}(\sigma_2) = \emptyset
  &
  \sigma = \sigma_1\cup\sigma_2
    }
\]
Whenever $\sigma$ is a valuation whose support contains the variables
of $v$, we write $\sigma(v)$ for the value where the variables of $v$
have been replaced with the corresponding values in $\sigma$, as follows:
\begin{itemize}
\item $\sigma(c_\alpha) = c_\alpha$,
\item 
 $\sigma(x) = v$ if $\{x\mapsto v\}\subseteq \sigma$,
\item 
 $\sigma(\inl{v}) = \inl{\sigma(v)}$,
\item 
 $\sigma(\inr{v}) = \inr{\sigma(v)}$,
\item 
  $\sigma(\pv{v_1}{v_2}) = \pv{\sigma(v_1)}{\sigma(v_2)}$.
\end{itemize}

\begin{definition}[Reduction]\label{def:reduction}
The reduction $\to$ is then defined as the smallest relation such that
$\isoterm\,t\to\isoterm\,t'$ whenever $t\to t'$ and such that,
provided that $\sigma[v_i] = v$, the redex
\[ \isobasique~v\] reduces to
$\sigma(v'_i)$. As usual, we write $s\to t$ to say that~$s$
rewrites in one step to $t$ and $s\to^*t$ to say that~$s$ rewrites to
$t$ in~0 or more steps.
\end{definition}

\subsection{Properties}

The language satisfies usual safety properties, and, although not
necessarily total, functions are indeed reversible. In this section we
formalize these results.

\begin{remark}
  The reduction is deterministic: if $s\to t$ and $s\to t'$ then
  $t=t'$.
\end{remark}

Because of the conditions set on patterns in the typing rule of
isos, the rewrite system is deterministic. Note that, since we do
not impose exhaustivity, the reduction might get stuck.
Nonetheless, the following properties hold~\cite{sabry2018symmetric}.

\begin{lemma}[Subject reduction]\label{lem:sr}
  If ~$\entailval s:a$ and $s\to t$ then $\entailval t:a$.\qed
\end{lemma}

\begin{lemma}[Termination]\label{lem:term}
  If ~$\entailval s:a$ then there exists a term $t$ that does not
  reduce such that $s\to^*t$.\qed
\end{lemma}

The small language presented in this section can be called
\emph{reversible} for the following reason.
Given an iso
\[\omega = \isobasique,\] we
can syntactically define the \emph{inverse} $\omega^{-1}$ as
the iso
\[\isobasiqueinv.\] This inverse
satisfies the following property.

\begin{lemma}[Inverse]\label{lem:inv}
  If $\entailiso \omega:a\iso b$ and $\entailval v : a$, and if
  $\omega\,v\to^* w$ with $w$ a value, then
  $\entailiso \omega^{-1}:b\iso a$ and we have that
  $\omega^{-1}\,w\to^* v$.\qed
\end{lemma}

\section{Inverse categories}
\label{sec:inv_cat}

The simple reversible language of Section~\ref{sec:language} can
easily be modeled within the category PInj of sets and partial
injective maps. Indeed, suppose that we write $\setden{a}$ for the set
of closed values of type $a$, that is, the set of values $v$ such that
$\entailval v:a$. Because of Lemmas~\ref{lem:sr} and~\ref{lem:term},
for each well-typed function $\entailiso \omega : a\iso b$ one can
define an injective set-map
$\setden{\omega} : \setden{a}\to\setden{b}$ as follows:
$\setden{\omega}(v) = w$ whenever $\omega\,v\to^*w$. Many of the
properties of PInj can be reflected in the language, for instance
partiality and union of functions with disjoint domain.

\begin{example}\label{ex:pattern}
  Let $\omega_1$ and $\omega_2$ be respectively
  defined as
  \begin{align*}
    &\entailiso\clauses{\clause{\inl{x}}{\inr{x}}}:a\oplus b\iso b\oplus
      a,\\
    &\entailiso\clauses{\clause{\inr{x}}{\inl{x}}}:a\oplus b\iso b\oplus
      a.
  \end{align*}
  They correspond to two partial set-functions $\setden{\omega_1}$ and
  $\setden{\omega_2}$ defined on the disjoint union
  $\setden{a}\uplus\setden{b}$: $\setden{\omega_1}$ is defined on the
  $\setden{a}$-component and $\setden{\omega_2}$ on the
  $\setden{b}$-component. 
  Defined as
  \[\clauses{\clause{\inl{x}}{\inr{x}}\clause{\inr{x}}{\inl{x}}},\]
  the iso $\omega$ of type $a\oplus b\iso b\oplus a$
  can
  be regarded as the \emph{union} of $\omega_1$ and $\omega_2$ ---and
  in the set-model, their denotation is such a union:
  $\setden{\omega}=\setden{\omega_1}\uplus\setden{\omega_2}$. This works
  only because the functions $\setden{\omega_1}$ and
  $\setden{\omega_2}$ are \emph{compatible} on their common domain
  of definition.
\end{example}

The category PInj has been extensively studied and its
structure analyzed within the framework of
\emph{inverse categories}~\cite{kastl1979inverse,cockett2002restriction-I,%
cockett2003restriction-II,%
cockett2007restriction-III,giles2014investigation,guo2012products}.
These categories
formalize the notion of \textit{partial inverse} morphisms and
also conveys a natural definition of joins (without relying on
coproducts) :---least upper bounds---,
which shall be shown as the best way to denote the
pattern-matching, as discussed in Example~\ref{ex:pattern}.

This section therefore aims at a rapid introduction to inverse
categories: it contains the necessary material needed to read and
understand the remaining sections: it is far from stating all the
results and interests of restriction and inverse categories. A reader
interested in the subject may for instance refer to~\cite{cockett2002restriction-I,%
cockett2003restriction-II,%
cockett2007restriction-III,giles2014investigation,guo2012products}
for further information.

\subsection{Restriction category}

The definition of a proper \emph{partial function} requires a formal
way to write the ``domain'' of a morphism $f$, through a \emph{partial
  identity} $\res f$ called \emph{restriction}.

\begin{definition}[Restriction~\cite{cockett2002restriction-I}]\label{def:restr}
  A restriction structure is an operator that maps each
  morphism $f:A\rightarrow B$ to a morphism $\res f:A\rightarrow A$
  such that
  \begin{align*}
    f \circ \res f &= f \quad\Label{eq:basic}
    & \res f \circ \res g &= \res g \circ \res f \quad\Label{eq:comm}
    & \res{f \circ \res g} &= \res f \circ \res g\quad\Label{eq:comp}
    & \res h \circ f &= f \circ \res{h \circ f} \quad\Label{eq:left}
  \end{align*}
  A morphism $f$ is said to be \textit{total} if $\res f = 1_A$.
  A category with a restriction structure is called a
  \textit{restriction category}.
\end{definition}

\begin{remark}
   When unambiguous, we write $gf$ for the composition $g\circ f$.
\end{remark}

\begin{example}
  Any category can be given a restriction structure by saying that all
  morphisms are total, but this is definitely not interesting as a
  model.  The standard non trivial example of restriction category is
  Pfn, the category of sets and partial functions. Given a partial
  function $f:A\rightarrow B$, we can define its restriction operator
  as $\res f(x) = x$ on the domain of $f$ and undefined otherwise.
\end{example}

Throughout this paper, we manipulate functors. These are often
required to keep the restriction structure intact; hence the following
definition.

\begin{definition}[Restriction functor~\cite{kaarsgaard2017join}]
  \label{def:restfun}
  A functor $F:\mathcal C \rightarrow \mathcal D$ is a
  \emph{restriction functor} if $\res{F(f)} = F(\res f)$ for all
    morphism $f$ of $\mathcal C$. The definition is canonically
    extended to bifunctors.
\end{definition}

\subsection{Inverse category}

Our goal is to denote reversible operations: this requires some form
of inverse.  The restriction operator gives a notion of domain of
morphisms, and moreover, $\res f$ is meant as an identity function on
this domain; thus the composition of $f$ with its presumed inverse
should be equal to its restriction. Hence the next
definition. Note that inverse categories were
  invented~\cite{kastl1979inverse} before restriction categories, but
  the order of definitions used here is believed more convenient by
  the authors.

\begin{definition}[Inverse category~\cite{kaarsgaard2017join}]
  \label{def:invcat}
  An \emph{inverse category} is a restriction category where all morphisms
  are partial isomorphisms; meaning that for $f:A\rightarrow B$, there
  exists a unique $f^{\circ} : B\rightarrow A$ such that
  $f^{\circ} \circ f = \res f$ and $f \circ f^{\circ} = \lc f$.
\end{definition}

The canonical example of inverse category is PInj, the category of
sets and partial injective functions. It is actually more than
canonical:

\begin{theorem}[\cite{kastl1979inverse}]\label{th:kastl}
  Every locally small inverse category is isomorphic to a subcategory
  of PInj.\qed
\end{theorem}

\begin{example}\label{ex:pid}
  Let us fix a non-empty set $S$.
  We then define PId$_S$ as the category with
  one object $*$ and with morphisms all of the sets $Y\subseteq S$. Composition is
  defined as intersection and the identity is $S$. Intersection also
  gives us a monoidal structure with unit $S$.  Note that the category
  PId$_S$ can be regarded as a subcategory of PInj, where each
  morphism $Y\subseteq S$ corresponds to a partial identity defined
  on $Y$.
  The category PId$_S$ can be endowed with a structure of inverse
  category by defining morphisms as their own restriction and partial
  inverse.
  
  From PId$_S$ we can define 
  PId$^{\oplus}_S$ as the inverse category whose
  objects are (generic) sets, and morphisms are defined as follows:
  $f:A\rightarrow B$ is a pair
  $f = \left( \sigma_f, \lbrace X^f_a\rbrace_{a\in
      \text{dom}(\sigma_f)} \right)$ where $\sigma_f$ is a partial
  injective map between sets $A\rightarrow B$, and where each $X^f_a$
  is a morphism of PId$_S$.
  Composition in PId$^{\oplus}_S$ is done pairwise with classical
  composition of functions for $\sigma$ and the composition in PId$_S$
  for the rest. The identity over $A$ is the pair made of the
  set-identity on $A$ together with $S$ for each $a\in A$.
  The restriction and partial inverse are generated from those of PInj
  and PId$_S$ for each element in the pair.
\end{example}

\subsection{Compatibility}
To give a denotation to isos in the programming language considered in
Section~\ref{sec:language}, it is compulsory to find a way to combine
terms such as $\omega_1$ and $\omega_2$ in Example~\ref{ex:pattern}.
More generally speaking, we need to combine morphisms of the same type
$A\to B$ to make
a ``join'' morphism.  First, we have to make sure that the morphisms
are \textit{compatible}: $f$ and $g$ are compatible if they are alike
on their common ``domain'', and can behave however they like when the
other does not apply.
Since we aim at building a model for (partial) bijections, compatibility
of the partial inverse is also checked.

\begin{definition}[Restriction compatible~\cite{kaarsgaard2017join}]
  \label{def:restcomp}
  Two morphisms $f,g:A\to B$ in a restriction category $\mathcal{C}$
  are restriction compatible if $f\res g = g\res f$.  The relation is
  written $f \smile g$.  If $\mathcal{C}$ is an inverse category, they
  are inverse compatible if $f\smile g$ and
  $f^{\circ} \smile g^{\circ}$, noted $f\asymp g$.
  A set $S$ of morphisms of the same type $A\to B$ is restriction compatible
  (\textit{resp.}  inverse compatible) if all elements of $S$ are
  pairwise restriction compatible (\textit{resp.} inverse compatible).
\end{definition}

\begin{example}\label{ex:compati}
  In PInj, let us consider
  $f,g:\lbrace a,b,c\rbrace\rightarrow\lbrace a,b,c\rbrace$ defined as
  identities on their domains where: $\text{dom}(f)=\{a,b\}$,
  $\text{dom}(g) = \{b,c\}$.  It is pretty clear that $f\asymp g$.
  However, if we consider
  $h:\lbrace a,b,c\rbrace\rightarrow\lbrace a,b,c\rbrace$ defined on
  $\{b,c\}$ as $h(b)=a$ and $h(c)=c$, then $f$ and
  $h$ are not compatible.
\end{example}

\subsection{Joins and ordering on morphisms}

When considering partial set-functions, a natural notion of order can
be built on functions by considering domain-inclusion: we can say that
$f\leq g$ if $g$ is defined on the whole ``domain'' of $f$ and both
behave the same there.
When considering models of reversible languages, joins are used to
model
pattern-matching~\cite{giles2014investigation,kaarsgaard2021join}:
each clause
is a partial function, and the complete pattern-matching can then be
represented with the \emph{join} of the clauses.
This notion can be extended to restriction categories as follows.

\begin{definition}[Partial order~\cite{cockett2002restriction-I}]\label{def:order}
  Let $f,g:A\to B$ be two morphisms in a restriction category. We
  then define $f \leq g$ as $g\res f = f$.
\end{definition}

\begin{example}
  Consider PId$_S$ from Example~\ref{ex:pid}: we have
  $X\leq Y$ iff $X\subseteq Y$.
  For PId$_S^\oplus$ we have $f \leq g$
  iff $\text{dom}(\sigma_f) \subseteq \text{dom}(\sigma_g)$ and, for
  all $a\in \text{dom}(\sigma_f)$, $\sigma_f(a) = \sigma_g(a)$, and,
  for all $a\in \text{dom}(\sigma_f)$, $X^f_a \subseteq X^g_a$.
\end{example}

\begin{property}\label{prop:order}
  Let us consider $f,g:A\rightarrow B$ such that $f\leq g$.
  Then, whenever $h:B\rightarrow C$, we have $hf\leq hg$. Similarly,
  whenever $h:C\rightarrow A$ we have $fh\leq gh$.\qed
\end{property}

As discussed in Example~\ref{ex:pattern}, the isos $\omega_1$ and
$\omega_2$ can be combine to form
the ``join'' morphism $\omega$. This notion of join is defined in
the context of restriction categories as follows.

\begin{definition}[Joins~\cite{guo2012products}]\label{def:join}
  A restriction category $\mathcal{C}$ is equipped with joins if for
  all restriction compatible sets $S$ of morphisms $A\to B$, there
  exists
  $\bigvee\limits_{s\in S} s:A\to B$ morphism of $\mathcal{C}$ such
  that, whenever $t : A \to B$ and whenever for all $s\in S$, $s\leq t$,
  \begin{align*}
    s &\leq \bigvee\limits_{s\in S} s ~~\Label{eq:joinleqright},
    & \bigvee\limits_{s\in S} s&\leq t~~\Label{eq:joinleqleft},
    &\res{\bigvee\limits_{s\in S} s}
    &=\bigvee\limits_{s\in S} \res s~~\Label{eq:joinres},
    & f\circ\left(\bigvee\limits_{s\in S} s\right)
    &= \bigvee\limits_{s\in
      S} fs~~ \Label{eq:joincompleft},
    & \left(\bigvee\limits_{s\in
      S} s\right)\circ g
    &=\bigvee\limits_{s\in S} sg
      ~~\Label{eq:joincompright}.
  \end{align*}
  Such a category is called a \emph{join restriction category}. An
  inverse category with joins is called a \emph{join inverse
  category}.
\end{definition}

\begin{example}
  Consider the morphisms $f,g$ from Example~\ref{ex:compati}. They are
  compatible, and 
  $f\vee g = 1_{\lbrace a,b,c \rbrace}$.
  It is easy to verify that
  \textit{e.g.} $f\leq 1_{\lbrace a,b,c \rbrace}$ and $g\leq 1_{\lbrace a,b,c \rbrace}$.
  Let us consider
  $h,k,l:\lbrace a,b,c \rbrace\rightarrow \lbrace a,b,c \rbrace$ such
  that $h(a) = b$ and undefined otherwise, $k(b) = c$ and undefined
  otherwise, and $l$ defined as $l(a)=b$, $l(b)=c$ and $l(c)=a$.
  We have $h\leq l$ and $k\leq l$. Besides,
  $h\vee k$ is the function whose domain is $\{a,b\}$ and sending
  $a\mapsto b$ and $b\mapsto c$.
  We therefore have $h\vee k \leq l$.
\end{example}

The join operator admits a unit, as follows.

\begin{definition}[Zero~\cite{kaarsgaard2017join}]\label{def:zero}
  Since $\emptyset \subseteq \text{Hom}_{\mathcal C} (A,B)$, and since
  all of its elements are restriction compatible, there exists a
  morphism $0_{A,B} \doteq\bigvee_{s\in\emptyset} s$, called
  \emph{zero}.
\end{definition}

\begin{lemma}[\cite{kaarsgaard2017join}]\label{lem:zeros}
  Whenever well-typed, the zero
  satisfies the following equations:
  $f0=0$,
  $0g=0$,
  $0_{A,B}^{\circ} = 0_{B,A}$,
  $\res{0_{A,B}} = 0_{A,A}$.\qed
\end{lemma}

\section{Semantics of isos}
\label{sec:sem_iso}

In this section, we turn to the question of representing the terms and
isos of the language presented in Section~\ref{sec:language}. Instead
of using the concrete category PInj as suggested in the header of
Section~\ref{sec:inv_cat}, we aim at using general, inverse
categories.

\subsection{Representing choice and pairing}

One of the problem to overcome is the denotation of pairing and
sums. If a standard categorical denotation for the former is a
monoidal structure, the latter is usually represented with
coproducts. Such a notion is however slightly insufficient. Indeed,
for being able to join independent clauses in pattern-matching (as
shown e.g. in Example~\ref{ex:pattern}), it has to interact well
with the restriction structure. This lead to the development of
\emph{disjointness tensor}. 
If Giles~\cite{giles2014investigation} was the first one to introduce
the notion of disjointness tensors, in this paper we rely on the
definition of~\cite{kaarsgaard2017join}.

\begin{definition}[Disjointness tensor~\cite{kaarsgaard2017join}]
  \label{def:disten}
  An inverse category $\mathcal{C}$ is said to have a
  \textit{disjointness tensor} if it is equipped with a restriction
  bifunctor
  $.\oplus .:\mathcal{C}\times\mathcal{C}\rightarrow\mathcal{C}$, with
  a unit object $0$ and morphisms $\iota_l:A\rightarrow A\oplus B$ and
  $\iota_r:B\rightarrow A\oplus B$ that are total and such that
  $\lc{\iota_l}~\lc{\iota_r} = 0_{A\oplus B}$.
\end{definition}


To model the types of Section~\ref{sec:language}, we therefore
essentially need a disjointness tensor and a monoidal
structure. Similar to what happen in the category PInj, they need to
relate through distributivity.

\begin{definition}[\cite{kaarsgaard2021join}]
  Let us consider a join inverse category equipped with a symmetric
  monoidal tensor product $(\otimes,1)$ and a disjointness tensor
  $(\oplus,0)$ that are join preserving, and such that there is an
  isomorphism
  $\delta_{A,B,C} : A \otimes (B\oplus C) \rightarrow (A\otimes
  B)\oplus (A\otimes C)$. This is called a \emph{join inverse rig
    category}.
\end{definition}

\begin{example}\label{ex:pidrig}
  The category PId$^{\oplus}_S$ is a join inverse rig category with
  the following structure.
  \begin{itemize}
  \item $A\otimes B = A\times B$, the usual cartesian product.
    If $f:A\rightarrow B$ and $g:C\rightarrow D$ then we define
      $\sigma_{f\otimes g}(a,c) \doteq (\sigma_f(a),\sigma_g(c))$ and 
      $X^{f\otimes g}_{(a,c)} \doteq X^f_a \cap X^g_c$.
    The unit $1$ is the singleton set.
  \item $A\oplus B = A\uplus B$, the disjoint union.
    If $f:A\rightarrow B$ and $g:C\rightarrow D$ we define
      $\sigma_{f\uplus g} (\iota_l(a)) \doteq \iota_l(\sigma_f(a))$,
      $\sigma_{f\uplus g} (\iota_r(c)) \doteq \iota_r(\sigma_g(c))$,
      $X^{f\uplus g}_{\iota_l(a)} \doteq X^f_a$,
      $X^{f\uplus g}_{\iota_r(c)} \doteq X^g_c$.
    The zero is the empty set.
  \end{itemize}
  Note how in $PId^\oplus_S$ if $S$ is of cardinality at least $2$
  there are several morphisms $1\to 1$. In particular a morphism
  $1\to1\oplus 1$ is not necessarily an injection.
\end{example}

\subsection{Orthogonality between morphisms}

In the description of the language, the definition of isos heavily
relies on orthogonality. In this section we define this notion
over morphisms in an inverse category.

General definitions of orthogonality in inverse categories exist, but
we have chosen to manipulate a more practical one. Our notion
introduced below verifies all the axioms set in
\cite{giles2014investigation} for orthogonality of morphisms.  We
shall then see in this section that orthogonal values produce
\textit{disjoint} morphisms.

\begin{definition}[Disjointness]\label{def:disj}
  $f:A\rightarrow B,\ g:A\rightarrow C$ are disjoint iff $\res f~\res g = 0$.
\end{definition}

To picture it, one may say that disjoint morphisms have an empty
common domain of definition.
We can then prove that some of the axioms of morphism
orthogonality stated in \cite{giles2014investigation} hold for
disjointness, expressed with the lemmas below.

\begin{lemma}[Left-composed disjointness]\label{lem:comportho}
  Let
  $f_1:A\rightarrow B_1,\ f_2:A\rightarrow B_2,\ g_1:B_1\rightarrow
  C_1,\ g_2:B_2\rightarrow C_2$ be morphisms.  If
  $\res{f_1}~\res{f_2} = 0$, then $\res{g_1 f_1}~\res{g_2 f_2} = 0$.\qed
\end{lemma}

\begin{lemma}[Right-composed disjointness]\label{lem:rightortho}
  Let $f:A\rightarrow B,\ g:A\rightarrow C,\ h:D\rightarrow A$ be
  morphisms.  If $\res f~\res g = 0$ then $\res{fh}~\res{gh} = 0$.\qed
\end{lemma}

\begin{lemma}[Disjoint compatibility]\label{lem:rescompa}
	If $\res{f_1}~\res{f_2} = 0$ then $f_1\smile f_2$.\qed
\end{lemma}

This last lemma is very natural: if two morphisms apply on strictly
different
domains, they are compatible.
It also allows to underline that disjointness
is a right choice to picture orthogonality.

\subsection{Semantics of types and values}

In order to model the types of the language presented in
Section~\ref{sec:language}, the most natural is to consider a
\emph{join} inverse \emph{rig} category to capture the tensor and the
sum-type.

Let us then fix such a join inverse rig category $\mathcal{C}$. Each
type $a$ is given an interpretation as an object in the category
$\mathcal{C}$, written $\intf a$.  If a type $\alpha$ has $n$
constants, it is necessary to choose $\intf\alpha$ such that there are
$n$ morphisms $f_\alpha^i: 1\rightarrow\intf\alpha$ ($i=1\ldots n$)
that are total and pairwise disjoint,
following Definition~\ref{def:disj}. The
set of these morphisms will be written $S_{\alpha}$.

A sequence of types or of terms is denoted with the tensor product
of the interpretations:
%
$\intf{\Delta} = \intf{a_1} \otimes \dots \otimes \intf{a_n}$
whenever
$\Delta \doteq x_1 : a_1, \dots, x_n : a_n $.
Then we define the semantics of values and terms by induction on their
definition: we set
%
$\intf{c^i_{\alpha}} = f^i_{\alpha}\in S_{\alpha}$.
The typing judgment of variables gives the following denotation:
%
$\intf{x:a \vdash x:a} \doteq 1_{\intf a} :\intf
a\rightarrow \intf a$.
Now let us consider $f = \intf{\Delta \vdash v : a}$. We have
the following denotation:
%
$\intf{\Delta \vdash
  \inl v : a\oplus b} \doteq \iota_l \circ f$,
and similarly for the right-projection.
If $f = \intf{\Delta_1 \vdash v_1 : a_1}$ and $g = \intf{\Delta_2
  \vdash v_2 : a_2}$, we can define
%
$\intf{\Delta_1, \Delta_2 \vdash \pv{v_1}{v_2} :
  a_1 \otimes a_2} \doteq f \otimes g$.
When $\Delta_1, \Delta_2$ are empty,
$f \otimes g:1\otimes 1\rightarrow \intf{a_1}\otimes\intf{a_2}$ will
be identified with the type
$1 \rightarrow \intf{a_1}\otimes\intf{a_2}$ since
there is a \textit{total} isomorphism between $1$ and $1\otimes
1$.

\begin{remark}
  By abuse of notation, we shall write $\intf v$ in place of
  $\intf{\Delta \vdash v : a}$ when the context is clear.
\end{remark}

The semantics of values $v_1,v_2$ of type $a$
within a context $\Delta$ are morphisms
$\intf{\Delta}\rightarrow \intf a$. The interesting part in terms of
orthogonality is the codomain of these morphisms: intuitively, if they
were sets, we would want them to be disjoint. This explains why we are
manipulating morphisms such as $\lcs v:\intf a\rightarrow \intf a$ in
the following lemma.

\begin{lemma}[Orthogonality]\label{lem:ortho}
  If $v_1 \perp v_2$, then $\lcs{v_1}~\lcs{v_2} = 0$.\qed
\end{lemma} 

Lemma~\ref{lem:ortho} is proven by induction on the definition of
orthogonality over values. It heavily relies on the fact that
$\lc{\iota_l}~\lc{\iota_r} = 0$, property of our category
$\mathcal{C}$ that we have settled with Definition~\ref{def:disten}.






\subsection{Iso semantics}


The aim of this section is to build the denotation of isos:
\[
  \entailiso\isobasique : a
  \iso b.
\]
Such a denotation should be a morphism $\intf a\rightarrow\intf b$. It
should directly depend on the denotation of the individual clauses, as
hinted in Example~\ref{ex:pattern}.
Within the inverse rig category $\mathcal{C}$, we aim at showing that
$\intf{\omega}\intf{v_i}$ is equal to
$\intf{v'_i}\intf{v_i}^\circ\intf{v_i}$.
This shows that the morphisms that we need to join are of the form
$\intf{v'_i}\circ\intf{v_i}^{\circ}$: we thus need them to be
compatible.  Their compatibility is a direct conclusion of
Lemmas~\ref{lem:comportho},~\ref{lem:rescompa} and~\ref{lem:ortho}.

\begin{lemma}\label{lem:compcompa}
  If $v_1 \perp v_2$ and $v'_1 \perp v'_2$, then
  $\intf{v'_1}\circ\intf{v_1}^{\circ} \asymp
  \intf{v'_2}\circ\intf{v_2}^{\circ}$.\qed
\end{lemma}

\begin{definition}\label{def:deniso}
  For a well-typed
  iso $\isobasique$,
  thanks to the orthogonality constraints on clauses the
  morphisms $\intf{v'_i}\circ\intf{v_i}^{\circ}$ form a family of
  pairwise compatible morphisms. We can rely on it to build the
  semantics of isos as
  \beq\label{eq:defisoden}
  \intf{\entailiso\clauses{\clause{v_1}{v'_1}\clause{v_2}{v'_2}~\ldots}
    : a \iso b} = \bigvee_{i} \intf{v'_i}\circ\intf{v_i}^{\circ}:
  \intf a \rightarrow \intf b. \eeq
  We can finally define the denotational semantics of the remaining
  terms of the language. Let $F$ be the denotation
  $\intf{\entailiso\omega:a\iso b}$ and $g$ be $\intf{\entailval t : a}$.
  Then $\intf{\entailval \omega t : b}$ is $F\circ g$.
\end{definition}

\subsection{Towards soundness}
\label{sec:towsound}
In order to validate the semantics, a standard expected result is
soundness: the fact that whenever $t\to t'$ we also have $\intf t =
\intf{t'}$. The main difficulty lies in the denotation
$\intf{\omega{}t}$ in Definition~\ref{def:deniso}. Indeed, suppose
that $\omega:a\iso b$ is defined as
$\clauses{\clause{v_1}{v'_1}\clause{v_2}{v'_2}}$.  Whenever a closed
term $t$ of type $a$ reduces to $w$, and that $\omega~w$ reduces,
thanks to the safety properties we know that there exists some $i$ and
a substitution $\sigma$ such that $\sigma[v_i]=w$ and $\omega{}t\to^*
\sigma(v'_i)$. We therefore need $\intf{\omega{}t}$ to be equal to
$\intf{\sigma(v'_i)}$, that is, since we should have $\intf t = \intf
w$ if we had soundness,
\beq\label{eq:denproppattern}
\left((\intf{v'_1}\circ\intf{v_1}^{\circ})\vee(\intf{v'_2}\circ\intf{v_2}^{\circ})\right)\circ\intf{w}
=
\intf{v'_i}\circ\intf{v_i}^{\circ}\circ \intf w.
\eeq
This amounts to ask that the pattern matching carries over joins in
the category.

In the literature~\cite{haghverdi2000phd,Hoshino12,%
giles2014investigation,axelsen2016join,kaarsgaard2021join}, the
problem is solved by capitalizing on sum-like monoidal tensors, in
particular the disjointness tensor of
Giles~\cite{giles2014investigation}. As sketched
in~\cite{kaarsgaard2021join}, one can follow the following
strategy. First, one has to make sure that the denotation maps all
types to objects of the form $1\oplus\cdots\oplus1$. Then,
capitalizing on the structure of language, one shows that the only
possible morphisms of type $1\to 1\oplus\cdots\oplus1$ representable
in the language are coproduct injections. The
behavior of the pattern-matching is then categorically mimicked,
getting Equation~\eqref{eq:denproppattern} directly from the properties of
the injections.

However, as for instance in the category PId$^\oplus_S$ of
Example~\ref{ex:pidrig}, in general join inverse rig categories
morphisms of type $1\to 1\oplus\cdots\oplus1$ are not necessarily
injections. This calls for a weaker condition, independent from the
coproduct. This is the subject of the next section.

\section{Pattern-matching categories}
\label{sec:pattern}

As discussed in Section~\ref{sec:towsound}, the denotation of iso is a
join over several morphisms, each mapping a particular pattern to its
image through the iso. In the operational semantics,
the ``choice'' of the pattern is expressed over the notion of
orthogonality of values: indeed, if a value matches a pattern, this
value is necessarily orthogonal to the other patterns. This is
ensured by the straightforward definition of orthogonality.
We have presented in Section~\ref{sec:towsound} a way to transfer this
notion of orthogonality to morphisms in join inverse rig categories, 
based on the structure of disjointness tensor.
%
%
%
In this section, we discuss a notion of categorical pattern-matching
independently from disjointness tensors, in particular generalizing
the approach of~\cite{giles2014investigation,kaarsgaard2021join}.

This is the main contribution of the paper.

\begin{example}\label{ex:nocopro}
  To illustrate our approach, consider a language only consisting of
  (1) tensors and (2) enum-types $\alpha=\{c_\alpha\}$ and
  $\beta=\{c^1_\beta,c^2_\beta,c^3_\beta\}$.
  Let us then consider the full subcategory of PInj, with objects
  consisting of the sets of cardinality power of $3$. This
  sub-category can serve as a sound model of the language, including
  the pattern-matching.
  Note how the handling of the pattern-matching is independent
  from any disjointness tensor as the category does not feature
  coproducts.
\end{example}

\subsection{Non decomposability and pattern-matching categories}

Concretely, in order to get pattern-matching, we manipulate
expressions of the form $(f\vee g) h$ and the idea would be to find a
property over $h$ which would lead to $(f\vee g) h$ being equal either
to $fh$ or to $gh$, without having to rely on coproduct
injections. For this matter, we introduce several definitions of
\emph{non decomposability} -- roughly meaning that the morphism cannot
be written with a join.

\begin{definition}[Strongly non decomposable]\label{def:snd}
  A morphism $h:A\rightarrow B$ in a join inverse category is called
  \textit{strongly non decomposable (snd)} when if $h=f\vee g$ with
  $f,g:A\rightarrow B$, then $f=0$ or $g=0$.
\end{definition}

\begin{definition}[Linearly non decomposable]\label{def:lnd}
  A morphism $h:A\rightarrow B$ in a join inverse category is called
  \textit{linearly non decomposable (lnd)} when if $f\leq h$ and
  $g\leq h$ with $f,g:A\rightarrow B$, then $f\leq g$ or $g\leq f$.
\end{definition}

\begin{definition}[Weakly non decomposable]\label{def:wnd}
  A morphism $h:A\rightarrow B$ in a join inverse category is called
  \textit{weakly non decomposable (wnd)} when if $h=f\vee g$ with
  $f,g:A\rightarrow B$, then $f\leq g$ or $g\leq f$.
\end{definition}

\begin{remark}
  Let us observe that $\text{snd} \Rightarrow \text{lnd} \Rightarrow
  \text{wnd}$.
\end{remark}

Even if they look alike, these definitions imply large differences. A
strongly non decomposable morphism does not have any morphism strictly
less than itself, except $0$. On the other hand, linear non decomposability
allows a certain order of morphisms, with a linear constraint. Finally the weak form gives very few information,
as shown in the following example.

\begin{example}[Weakly non decomposable morphism]
  \label{ex:nwc0}
  Consider a subcategory of PInj with one object $\{a,b,c\}$ and with
  the identity and the four partial isos: $f$ of domain $\{a,b\}$, $g$
  of domain $\{a\}$, $h$ of domain $\{b\}$ and $0$ of empty domain.
  This subcategory is a join inverse category.
  In this subcategory, the identity is weakly non decomposable: it
  cannot be written as a join without itself in it. Nonetheless, the
  morphism $f$ is decomposable: the property of weakly non
  decomposability is not downwards closed, contrary to the two other
  notions. Weak non-decomposability then lead to very different
  results from the other notions.
\end{example}



\begin{definition}[Pattern-matching]\label{def:match}
  A weakly (resp. linearly, strongly) pattern-matching category
  $\mathcal C$ is a join inverse category in which for all
  $f,g \in \text{Hom}_{\mathcal C} (A,B)$ and
  $h \in\text{Hom}_{\mathcal C} (C,A)$, if $f\smile g$ and $h$ is
  weakly (resp. linearly, strongly) non decomposable, then
  \[(f\vee g) h = fh\quad\text{ or }\quad(f\vee g) h = gh.\] Note that we do
    not need the compatibility of the inverses to define the
    pattern-matching.
\end{definition}

This straightforward definition of what we would like to have in our
category is, however, difficult to grasp and manipulate; therefore we
introduce an equivalent presentation: the correspondence 
between the two definitions is shown in Theorem~\ref{th:thecorresp}.

\begin{definition}[Consistency]\label{def:consis}
  A join inverse category is said weakly (resp. linearly, strongly)
  \textit{consistent} when $h:A\rightarrow B$ weakly (resp. linearly,
  strongly) non decomposable implies that for all morphism
  $f:B\rightarrow C$, the morphism $fh$
  is weakly (resp. linearly, strongly) non
  decomposable.
\end{definition}

\begin{theorem}\label{th:thecorresp}
  A weakly (resp. linearly, strongly) consistent join inverse category
  is a weakly (resp. linearly, strongly) pattern-matching category.\qed
\end{theorem}

Indeed, if a join $f\vee g$ is composed on the left with a non
decomposable morphism $h$, consistency ensures that $fh\vee gh$ stays non
decomposable. This means that it can be only be one of the morphisms
that form the join, which is exactly pattern-matching.

\begin{theorem}\label{th:wconsis}
  A weakly pattern-matching category is weakly consistent.\qed
\end{theorem}

This theorem validates the choices of definition made above, and it
underlines a strong link between pattern-matching and non
decomposability.

Notwithstanding, an inverse category is not necessarily a weakly
pattern-matching category:

\begin{example}[Not weakly consistent]\label{ex:nwc}
  Consider the subcategory of PInj presented in Example~\ref{ex:nwc0}.
  Although $\text{id}_{\{a,b,c\}}$ is weakly non decomposable, the
  composition $(g\vee h) \text{id}_{\{a,b,c\}}$ is equal to
  $g\vee h = f$, different from both $g$ and $h$: the subcategory is
  not weakly consistent.
\end{example}


Thus, if we were willing to work with weak non decomposability, not every
inverse category can be involved. As shown below, the notions of linear
and strong non decomposability let us consider any inverse category.
Of course, the complexity does not disappear: it simply moves from
the definition of the category to the number of morphisms that we can
consider, since linear and strong non decomposability are far more
demanding properties than the weak one.


\begin{theorem}\label{th:sconsis}
  A join inverse category is strongly consistent.\qed
\end{theorem}

Even if it is not a surprise, this result is interesting. Strongly non
decomposability is a very demanding property; but since it implies
some morphisms being equal to $0$, composing on the left naturally
does not change the property.

\begin{example}\label{ex:lndnsnd}
  A morphism in a join inverse category can be linearly non
  decomposable without being strongly non decomposable. Indeed,
  consider the (join inverse) sub-category of PInj whose only object
  is $\{a,b\}$, and with morphisms $0$, $\text{id}$ and $f$, the
  partial identity whose domain is $\{a\}$. Then we have
  $0<f<\text{id}_{\{a,b\}}$. While it is linearly non-decomposable,
  the identity is not strongly non-decomposable.
\end{example}

Fortunately, we still have the following result:

\begin{theorem}\label{th:lc}
  A join inverse category is linearly consistent.\qed
\end{theorem}

\begin{remark}
  Linear decomposability is then the best property to consider for
  pattern-matching.
  In the context of pattern-matching categories, linear non
  decomposable maps then essentially correspond to ``pure'' values
  waiting to be matched against a set of ``orthogonal'' clauses
  ---this orthogonality being captured by the compatibility of the
  morphisms representing the clauses.
\end{remark}

\subsection{The case of our language}

The previous subsection being an attempt to develop a categorical
theory of pattern-matching, we can now take a look at how it would
apply to the case of the minimal language of
Section~\ref{sec:language}, whose only base type is the
$\mathtt{unit}$ type with one single constant. Its semantics in a join
inverse rig category $\mathcal{C}$ is straightforwardly given by the
identity $\text{id}_1$ over $1$, the unit of the monoidal structure.
We are in the context of Section~\ref{sec:towsound}, where all closed
values are denoted with morphisms $1\to1\oplus\cdots\oplus1$, but
where in the category, in general morphisms of this type are
not necessarily injections ---and there are more than one morphism $1\to1$.

\begin{theorem}[Denotation of closed values]\label{lem:denct}
  In a join inverse rig category $\mathcal{C}$, if $\text{id}_1$ is
  weakly (resp. linearly, strongly) non decomposable, then the
  denotation of a closed value $v$ of the minimal language is weakly
  (resp. linearly, strongly) non decomposable.\qed
\end{theorem}

Then in any join inverse category with $\text{id}_1$ linearly
---or strongly--- non decomposable, in the context of
Sec.~\ref{sec:sem_iso} the previous results allow us
to apply pattern-matching:
$\left(\bigvee_{i} \intf{v'_i}\intf{v_i}^{\circ}\right) \intf v =
\intf{v'_{i_0}}\intf{v_{i_0}}^{\circ}\intf v$.

\begin{example}\label{ex:pidsub}
  Example~\ref{ex:pidrig} is a nice way to picture a $\text{id}_1$
  that is non-trivial. Suppose that $S$ contains at least two
  elements. Then $\text{id}_1$ in PId$^{\oplus}_S$ is not even weakly
  non decomposable; if we consider the subcategory
  SubPId$^{\oplus}_{\{a,b,c\}}$ with $S=\lbrace a,b,c \rbrace$, and
  with corresponding morphisms on PId$_S$ $\lbrace a\rbrace$,
  $\lbrace a,b\rbrace$ and $\{a,b,c\}$ only, then $\text{id}_1$ is
  linearly non decomposable but not strongly so.

  Note how
  SubPId$^{\oplus}_{\{a,b,c\}}$ is indeed a join inverse rig
  category, as the tensor is derived from the intersection on
  sets.
\end{example}


\subsection{Soundness and adequacy}
\label{sec:adequacy}

From Theorem~\ref{lem:denct} we can derive the fact that any join
inverse rig category where $\text{id}_1$ is linearly non-decomposable
forms a sound and adequate model for the minimal language. The
remaining results assume such an ambient semantics.

\begin{lemma}[Substitution]\label{lem:pmsound}
  \label{eq:sound}
  In the minimal language, for any value $\Delta \entailval v_{i_0}:a$,
  $\Delta\entailval v'_{i_0}:b$ and $\entailval v: a$, we have
  $\intf{\sigma(v'_{i_0})}
  =\intf{v'_{i_0}}\intf{v_{i_0}}^{\circ}\intf v$.\qed
\end{lemma}

\begin{corollary}[Soundness]\label{cor:directsound}
	If $t\rightarrow t'$, then $\intf t = \intf{t'}$.\qed
\end{corollary}

\begin{definition}[Operational equivalence]\label{def:opeq}
  Let's consider two terms $t_1,t_2$. We define $t_1 \equiv_{op} t_2$ as
  both $t_1\rightarrow^* v$ and $t_2\rightarrow^* v$.
\end{definition}

\begin{theorem}[Adequacy]\label{th:adeq}
  If $t_1$ and $t_2$ are closed, well-typed terms, then
  $t_1 \equiv_{op} t_2$ iff $\intf{t_1} =
  \intf{t_2}$.\qed
\end{theorem}

\section{Discussion and conclusion}

Consider again Example~\ref{ex:pidsub}. Following the same technique
presented in the literature~\cite{kaarsgaard2021join}, one could be
able extract from the minimal language of Section~\ref{sec:language}
a sub-category of PId$^\oplus_{\{a,b,c\}}$ able to capture a notion
of pattern-matching. However, this approach does not tell us
anything about the categorical structures required to capture it:
the machinery presented in Section~\ref{sec:pattern} aims at
answering this question.

In particular, we discussed in Example~\ref{ex:pidsub} a subcategory
of PId$^\oplus_{\{a,b,c\}}$: in SubPId$^{\oplus}_{\{a,b,c\}}$, the
morphism id$_1$ is linearly non-decomposable; Lemma~\ref{lem:pmsound}
then shows that Equation~\eqref{eq:denproppattern} of
Section~\ref{sec:towsound} is correct in SubPId$^{\oplus}_{\{a,b,c\}}$
while Theorem~\ref{th:adeq} guarantees that this subcategory is
actually adequate: it strictly sits in between the image of the
language and the whole category PId$^\oplus_{\{a,b,c\}}$.

In general, although the notions of pattern-matching and linear
non-decomposability we presented in this paper are sufficient to
recover a notion of pattern-matching, a question that is however left
open is whether they are necessary. This is left as future work.

\section*{Acknowledgments}

This work was supported in part by the French National
Research Agency (ANR) under the research project SoftQPRO
ANR-17-CE25-0009-02, by the DGE of the French Ministry of
Industry under the research project PIA-GDN/QuantEx
P163746-484124 and by the STIC-AmSud project Qapla' 21-SITC-10.

\bibliographystyle{eptcs}
\bibliography{ref}





\end{document}